\documentclass[prx,floatfix,showpacs,superscriptaddress,longbibliography,twocolumn]{revtex4-1}

\usepackage[utf8x]{inputenc}

\usepackage[USenglish]{babel}

\usepackage{amsmath,amssymb,amsfonts}

\usepackage{bm}

\usepackage{epsfig}
\usepackage{subfigure}

\usepackage[colorlinks=true,linkcolor=blue,citecolor=red]{hyperref}

\newcommand{\equ}[1]
{Eq.~(\ref{#1})}

\newcommand{\figu}[1]
{Fig.~\ref{#1}}

\newcommand{\secu}[1]
{Sec.~\ref{#1}}

\def\=={\equiv}

\def\cG0{{\cal G}_0} 
\def\cG{{\cal G}}

\def\=={\equiv}

\newcommand{\be}{\begin{equation}}
\newcommand{\ee}{\end{equation}}

\begin{document}

\date{\today}

\author{Francesco~Grandi}
\affiliation{Department of Physics, University of Erlangen-N\"urnberg,
  91058 Erlangen, Germany}

\author{Martin~Eckstein}
\affiliation{Department of Physics, University of Erlangen-N\"urnberg,
  91058 Erlangen, Germany}

\title{Ultrafast metal-to-insulator switching in a strongly correlated system}

\begin{abstract}

Light-manipulation of correlated electronic phases in solids offers the tantalizing prospect of realizing electronic devices operating at the ultrafast time-scale. In this context, the experimental realization of non-equilibrium transitions from a metal to a band or Mott insulator has shown to be particularly elusive. Using dynamical mean-field theory, we study a simple model representing the main physical properties of the oxygen-enriched compound LaTiO$_{3+x}$. By properly optimizing the photo-doping of electrons from a low-energy band into the valence states of the system, we show it is possible to induce a valence transition from a correlated metallic state to a Mott insulator at ultrashort time scales and to contain the heating during this process, with the final non-thermal valence insulator having almost the same effective temperature of the starting metal.
\end{abstract}
\maketitle

\section{Introduction}\label{secI}

Transition metal compounds (TMC) are an intriguing class of materials in condensed matter physics due to the conspiracy of charge, spin, orbital and lattice degrees of freedom \cite{khomskii_2014}. Many representatives of this group show a metal-to-insulator transition upon changing intensive variables such as the temperature $T$, the external pressure or the chemical potential \cite{PhysRevLett.3.34,Imada1998RMP}. The equilibrium insulator-to-metal phase transition is typically realized upon increasing $T$ if it involves some form of symmetry breaking (lattice dimerization, orbital order, antiferromagnetism, etc.) \cite{PhysRevB.45.8209,McWhan1973PRB,nature12425,Frandsen2016}. On the other hand, if both metal and insulator have the same symmetry, the transition to the insulator might occur with increasing instead of decreasing $T$, simply because the entropy of the Mott state can be higher than that of the metallic state. For example, after locking the charge, the remaining spin degree of freedom in a spin-disordered Mott insulator implies a high entropy $S_{\text{Mott}} \sim \ln \left( 2 \right)$ per lattice site \cite{RevModPhys.68.13}. In contrast, the entropy in a Fermi liquid vanishes for low temperatures like $S_{FL} \left(T\right) \propto T$ so that the transition to a Mott state can be driven by a temperature increase if the spin order remains frustrated.

While some of the equilibrium properties of these materials are still not fully understood, a large interest in their non-equilibrium behavior has grown recently \cite{giannetti2016,Basov2017}. In particular, the tantalizing prospect of realizing a Mottronic switch between insulating and metallic states has attracted much attention \cite{Guiot2013,PhysRevLett.117.176401,LiNano2017,ronchi2020nonthermal,delValle2019,Janod2015}. Even if ultrafast insulator-to-metal phase transitions have been reported in several materials  \cite{PhysRevB.70.161102,Stojchevska177,Wolf-PRL2014,nat_comm_V2O3,PhysRevApplied.11.014054}, the opposite phase change is harder to achieve. Since a laser excitation typically increases the effective temperature and the total entropy of a system, it is generally easier to move from a symmetry broken insulating state to a metallic one. Nevertheless, one can imagine several pathways to laser-induce the opposite transition from metallic to insulating behavior if the final insulating state is of the Mott kind. Firstly, with the entropic argument given above, the metal-to-insulator phase transition may be induced in some specific cases simply by a laser-induced heating of the electrons. An alternative scenario is that a photo-induced transfer of charges from other bands into a previously doped valence band can bring the latter to commensurate filling. While this will not make the system insulating as a whole (because holes are generated in other previously filled  bands), Mott physics can arise in the valence band. In the following, we will term this transient state a {\em valence Mott insulator}. Again, the main difficulty is that a rapid charge transfer typically leads to a strong increase of the entropy, working against correlation effects. The question is, therefore, whether suitably optimized protocols can be designed to generate such a valence Mott state within short times.

Even if the entropy of the system as a whole is expected to grow after the pulse, the entropy might transiently decrease in a given subsystem, leaving the remainder at high entropy. This concept has been explored theoretically during the past years \cite{PhysRevLett.114.137001,Nature_cooling,PhysRevLett.120.220601,PhysRevB.102.241103} as a pathway to induce or enhance electronic orders. Recently, it has been observed that also a charge transfer between different bands can realize this entropy redistribution \cite{Nature_comm_cooling}. By transferring (``evaporating'') holes from  the valence band to localized core levels, entropy is reshuffled between the valence and the core states, with the latter having practically zero entropy before the arrival of the pulse. In \cite{Nature_comm_cooling}, a single-band strongly correlated metal is considered as the initial state. By photo-doping from a narrow core level, the occupation of the valence band becomes almost commensurate with the lattice, and the effective temperature of the valence subsystem can decrease and eventually reach values below the equilibrium antiferromagnetic Ne\'el temperature. This mechanism has therefore been termed \textit{cooling by photo-doping}. With a similar mechanism, an enhancement of the $\ensuremath{\eta}$-paired superconducting susceptibility in the half-filled single-band Hubbard model has been predicted \cite{PhysRevB.102.165136}.

In this paper, we aim to investigate the cooling by photo-doping procedure in a multi-band setup. While the simulations will be based on a simplified two-band model in order to make them feasible, this model is designed to capture the important properties of the $3$d$^{1}$ perovskite series. The latter includes the metallic compounds BaVO$_{3}$, SrVO$_{3}$, and CaVO$_{3}$, and the Mott insulators LaTiO$_{3}$ and YTiO$_{3}$ \cite{NISHIMURA2014710,BANNIKOV2016119,Pavarini_2005,PhysRevLett.92.176403}. In all these materials, the transition metal atoms sit at the centre of a corner-sharing O$_{6}$ octahedron, while the cations occupy the interstitial space between the octahedra. The cubic crystal field splits the $3$d manifold in two higher energy $\text{e}_{\text{g}}$ and three lower energy $\text{t}_{2 \text{g}}$ levels, filled with a single electron. We particularly focus on LaTiO$_{3}$, the first insulating member of the series. In this compound, a distortion of the GdFeO$_{3}$ kind lowers the lattice symmetry to orthorhombic and further splits the $\text{t}_{2 \text{g}}$ states into a lower energy $\text{a}_{\text{g}}$ level and a higher energy $\text{e}_{\text{g}}'$ doublet \cite{PhysRevB.77.115350}. The material has a Mott-gap of $\sim 0.2 \ \mathtt{eV}$, which is small compared to the bandwidth of approximately $W \sim 2.1 \ \mathtt{eV}$ \cite{PhysRevB.48.17006,PhysRevB.51.9581}. A broad band, mainly derived from oxygen p-orbitals, lies well below the $\text{t}_{2 \text{g}}$ manifold \cite{PhysRevB.87.035107}. Below temperature $T_{N} \sim 146 \ \mathtt{K}$, the system undergoes a transition from a paramagnet to G-type antiferromagnetism \cite{PhysRevB.75.224402}.

The small value of the energy gap observed in this material suggests its proximity to a metal-insulator phase boundary, making it interesting for Mottronics \cite{PhysRevB.70.245125}. Moreover, this compound shows a doping-driven phase transition by growing it under oxygen atmosphere so that the high-temperature non-stoichiometric compound LaTiO$_{3+\text{x}}$, with $x \sim 0.04$, is a correlated metal with well defined quasi-particle properties \cite{doi:10.1002/adma.201706708,PhysRevB.51.9581,PhysRevB.59.7917,Lichtenberg1991}. In the following, we explore the possibility to drive the d-valence bands in the oxygen-enriched compound LaTiO$_{3+\text{x}}$ from the metallic to the Mott insulating state by photo-doping electrons into the energy states around the Fermi level from the lower-energy p-derived band. In other words, the oxygen p-bands are supposed to take the role of the core bands in the cooling by photo-doping scenario, in spite of these bands being certainly not localized, but in contrast even broader than the valence bands. Depending on the form of the excitation pulse, we indeed find that the valence band state can be turned into a Mott insulator with an effective temperature almost equal to the one of the starting metal on ultrafast timescales due to an entropy increase in the oxygen bands, similar to the cooling by photo-doping protocol.

The article is organized as follows. In \secu{secII}, we define a minimal model for the description of the photo-doping process in LaTiO$_{3+\text{x}}$ (\secu{secIIA}), discuss the excitation protocol (\secu{secIIB}) and the solution of the model using non-equilibrium dynamical mean-field theory (DMFT) (\secu{secIIC}). In \secu{secIII}, we present our results, showing the properties of the charge transfer process (\secu{secIIIA}) and the characteristics of the final state for several pulse lengths and optimized pulse parameters (\secu{secIIIB}). Finally, \secu{secIV} is devoted to concluding remarks.

\section{Model and Method}\label{secII}

\subsection{Two-band model}\label{secIIA}

This section aims to define a minimal model that can describe the photo-induced metal-to-valence insulator transition in the oxygen-enriched strongly correlated compound LaTiO$_{3+x}$. As we already commented, the parent compound LaTiO$_{3}$ has the three $\text{t}_{2 \text{g}}$-derived states (the lower energy $\text{a}_{\text{g}}$ level and the $\text{e}_{\text{g}}'$ doublet) at the valence filled with a single electron. Instead of three bands we consider two, where the degeneracy between the states is lifted by a crystal field interaction that mimics the effect of the GdFeO$_{3}$-like distortion. This way, we neglect the double degeneracy of the $\text{e}_{\text{g}}'$ states, which should however not alter the qualitative picture we aim to present because these states are mostly empty. Indeed, having two bands is already enough to explore the role of orbital imbalance in this system and to describe the nearly inter-band gap predicted in several theoretical analysis \cite{PhysRevB.77.115350,Pavarini_2005}. In our two-band model, the undoped Mott insulator corresponds to  quarter filling (one electron per site), and the total filling per site will be fixed to $0.92$ for the doped compound. In LaTiO$_{3+\text{x}}$, a valence occupation  $0.92$ would correspond to oxygen-doping $x=0.04$, enough for the material to show well-defined correlated metal properties. 

The Hamiltonian of the two-band model is given by the sum of a local ($H_{\text{loc}}$) and a non-local ($H_{\text{hop}}$) contribution, 
\be \label{tot_ham}
H = H_{\text{hop}} + H_{\text{loc}} \;.
\ee
The local part is given by
\be \label{loc_ham}
\begin{split}
H_{\text{loc}} = & -\frac{\Delta}{2} \sum_{\boldsymbol{R}} \left( n_{\boldsymbol{R},1} - n_{\boldsymbol{R},2} \right) - \mu \sum_{\boldsymbol{R}} n_{\boldsymbol{R}} \\
& + U \sum_{\boldsymbol{R},a} n_{\boldsymbol{R},a,\uparrow} n_{\boldsymbol{R},a,\downarrow} + U' \sum_{\boldsymbol{R}} n_{\boldsymbol{R},1} n_{\boldsymbol{R},2} \;, 
\end{split}
\ee
where the index $a=1,2$ labels orbitals; although this is a simplified model, we will refer to orbital $1$  and $2$ as $ \text{a}_{\text{g}}$ and $\text{e}_{\text{g}}'$, respectively. The spin takes values  $\sigma = \uparrow, \downarrow$; $n_{\boldsymbol{R}} = \sum_{a,\sigma} n_{\boldsymbol{R},a,\sigma}$ is the occupation on site $\boldsymbol{R}$, $\Delta$ is the crystal field splitting coupled to the orbital imbalance $\tau^{z}_{\boldsymbol{R}} = \frac{1}{2} \left( n_{\boldsymbol{R},1} - n_{\boldsymbol{R},2} \right)$, $U$ is the intra-band Hubbard interaction, and $U'$ is the  inter-band Hubbard repulsion. The Hund's coupling is relevant only for doubly occupied states and, moreover, is expected to be small compared to the bandwidth \cite{PhysRevResearch.2.013298}, so its contribution is neglected. The chemical potential $\mu$  is used to fix the average occupation $\langle n_{\boldsymbol{R}} \rangle=0.92$ on each site. 

The hopping part in Eq.~\eqref{tot_ham} is written as
\be \label{hop_ham}
H_{\text{hop}} = - \sum_{\langle \boldsymbol{R},\boldsymbol{R'} \rangle, \sigma} \sum_{a,a'} v_{\boldsymbol{R}, a;\boldsymbol{R'}, a'} \left( c^{\dagger}_{\boldsymbol{R},a,\sigma} c_{\boldsymbol{R'},a',\sigma} + \text{H.c.} \right) \;.
\ee
The operator $c_{\boldsymbol{R},a,\sigma}$ ($c^{\dagger}_{\boldsymbol{R},a,\sigma}$) annihilates (creates) an electron on site $\boldsymbol{R}$ in orbital $a$ with spin $\sigma$, and the summation extends just over nearest-neighbor sites. The hopping matrix $v_{\boldsymbol{R}, a;\boldsymbol{R'}, a'}$ allows both intra- and inter-band charge transfers, so that the occupation of each band is not individually conserved. We take the intra-band hopping $v_{\boldsymbol{R}, a; \boldsymbol{R'}, a} = v/\sqrt{z}$ equal for all nearest neighbors ($z$ is the coordination number). The inter-band hopping $v_{\boldsymbol{R}, 1; \boldsymbol{R'},2}=v_{\boldsymbol{R}, 2;\boldsymbol{R'},1}$ has absolute value $v'/\sqrt{z}$ and phases such that 
\be \label{symmetry}
\sum_{\boldsymbol{R}'}  v_{\boldsymbol{R}, 1;\boldsymbol{R'}, 2} =0 \;,
\ee
so that the two bands do not mix at the $\Gamma$-point. Below, we will solve the model on an infinitely-coordinated Bethe lattice, mainly because this implies a closed form of the DMFT self-consistency (as described in Sec.~\ref{secIIC}) and therefore allows to perform simulations up to longer times. It can be expected that the simulations on the Bethe lattice give qualitatively the same physics as on a cubic lattice if the bandwidth and energy gaps are properly matched and if the local environment [Eq.~\eqref{symmetry}] of every site is correctly represented.

The parameters in \equ{tot_ham} are chosen so that the bandwidth and energy gaps are comparable to the LaTiO$_{3}$-derived compound: the energy scale is set to the half-bandwidth, which is $2 v$, and time is measured in units of $\hbar / \left( 2 v \right)$. Because the bandwidth in LaTiO$_{3}$ is $2.1 \ \mathtt{eV}$ \cite{PhysRevB.77.115350,Pavarini_2005}, we can almost read the energy units in our plots in electronvolts. Further, we fix the crystal field to $\Delta = 0.15$ \cite{PhysRevLett.94.056401,PhysRevLett.92.176403,PhysRevB.94.245109} and the inter-band hopping $v' = 0.15$ \cite{Pavarini_2005}. Finally, we set the values of the intra- and inter-orbital Hubbard repulsion to $U = 3.10$ and $U' = 2.294$, respectively, to reproduce experimentally obtained energy gap of the quarter-filled (undoped) system, as well as the inter-band nature of the gap. This way, the ratio $U'/U$ has the same value as obtained in more sophisticated calculations \cite{PhysRevLett.94.056401}. The initial inverse temperature of the problem is $\beta = 10$, well above the antiferromagnetic critical temperature. 

\begin{figure}
 \centering \includegraphics[width=0.5\textwidth]{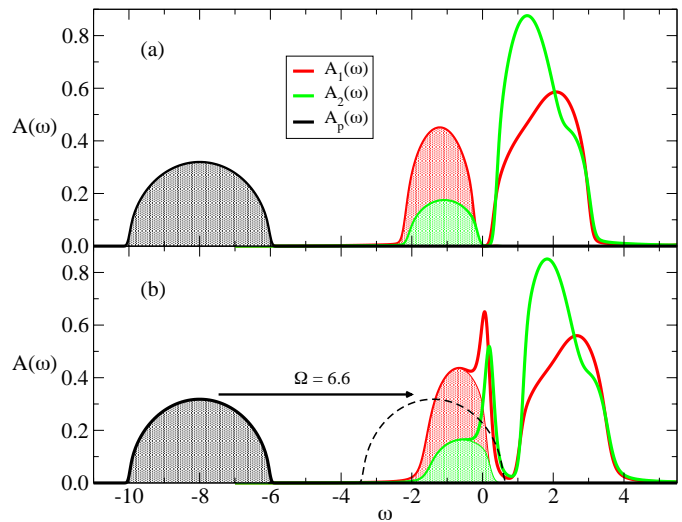}
  \caption{ 
  Spectral functions for the two orbitals $A_{1} \left( \omega \right)$ (red) and $A_{2} \left( \omega \right)$ (green), and  the non-interacting $p$-band $A_{p} \left( \omega \right)$ (black), together with their occupations (shaded areas) for the undoped (panel (a)) and the doped (panel (b)) systems, where we consider an average filling per site $n=1$ and $n=0.92$, respectively. The effect of a finite hopping frequency $\Omega$ is to shift the fermionic bath up in energy, as schematically shown by the arrow and the dashed line  in panel (b). The frequencies $\omega$ are shown with respect to the Fermi level of the system. The values of the double occupancies and of the orbital imbalance of the systems are given by $d \sim 0.0300$ and $\tau_{z} \sim 0.2197$ for $n =1$, and by $d \sim 0.0234$ and $\tau_{z} \sim 0.2077$ for $n = 0.92$.}
  \label{fig_spectr_func_vs_om_beta_10_bath}
\end{figure}

Figure~\ref{fig_spectr_func_vs_om_beta_10_bath}(a) shows the orbitally-resolved spectral functions of the model obtained using DMFT (Sec.~\ref{secIIC}) at $\beta =10$. The system features a lower band of mostly $a_{g}$-character, a Hubbard gap of about $0.2$, and an upper Hubbard band (UHB) which is of predominantly $e_{g}'$ character at the lower edge. Upon hole-doping (Fig.~\ref{fig_spectr_func_vs_om_beta_10_bath}b), a quasiparticle band is formed at the lower Hubbard band edge. In addition to the $d$-bands, we add an additional low-lying noninteracting band with semi-elliptic density of states, representing the $p$-band manifold (depicted by $A_p(\omega)$ in Fig.~\ref{fig_spectr_func_vs_om_beta_10_bath}). 

\subsection{Excitation pulse}\label{secIIB}

The system is excited with a time-dependent hybridization $v_{pd}(t)=AE(t)$ between the $d$ and $p$ bands, representing a dipolar transition driven by the electric field of a laser $E(t)$, with the dipolar matrix element $A$; choosing $A=1$ in the following sets the unit of the electric field. As in Ref.~\cite{Nature_comm_cooling}, we will investigate the use of chirped laser pulses \cite{STRICKLAND1985219} for controlling the system.  The functional form for the electric field is taken as
\be \label{pulse}
E(t) = E_0 f_{\text{env}}(t) \text{sin} \left(\Omega(t) t \right),
\ee
characterized by  the amplitude $E_0$, a linearly chirped frequency
\be 
\Omega \left( t \right) = \Omega_{\text{st}} + \frac{\Omega_{\text{end}} - \Omega_{\text{st}}}{t_{\text{pulse}}}  t 
\ee
with pulse duration $t_{\text{pulse}}$ and start (end) frequency $\Omega_{\text{st}}$ ($\Omega_{\text{end}}$), and a smooth envelope of duration $t_{\text{pulse}}$, $f_{\text{env}}(t)= \left[( 1 + e^{\alpha (t - t_{\text{pulse}}) + 6} ) ( 1 + e^{- \alpha t + 6} )\right]^{-1}$. In the following, we consider pulses with different durations $t_{\text{pulse}}$, and optimize all parameters $\alpha$, $E_0$, $\Omega_{\text{st}}$ and $\Omega_{\text{end}}$ to achieve certain final states (Sec.~\ref{secIII}).

\subsection{DMFT solution}\label{secIIC}
We solve the model using nonequilibrium DMFT \cite{RevModPhys.68.13,Aoki2014,SCHULER2020107484}. In DMFT, the original Hamiltonian \equ{tot_ham} is mapped onto a single-site impurity with local properties that correspond to \equ{loc_ham}. The impurity is hybridized with a bath characterized by a self-consistent hybridization matrix $\Delta_{a,b} \left( t,t' \right)$, with orbital indexes $a$, $b$. Because we consider relatively high temperatures and interaction strengths, we can to first approximation apply the non-crossing approximation (NCA) as impurity solver \cite{PhysRevB.82.115115}. Our choice of the hopping matrix in \equ{hop_ham}, and in particular the condition in \equ{symmetry}, allows to solve the problem entirely in terms of the diagonal components $\Delta_{a} \left( t,t' \right) = \Delta_{a,a} \left( t,t' \right)$ of the hybridization matrix. On the Bethe lattice, the self-consistency relation takes a closed form,
\be
\Delta_{a} \left( t,t' \right) = v^{2} G_{a} \left( t,t' \right)  + v'^{2} G_{\bar{a}} \left( t,t' \right) + \Delta_{p} \left( t,t' \right)  \;,
\ee
where $G_{a} \left( t,t' \right)$ is the local contour-ordered electronic Green's function for band $a$, $\bar{a}=2$ for $a=1$ and vice versa, and $\Delta_{\text{p}} \left( t,t' \right)$ is the hybridization function due to the coupling of the valence bands to the non-interacting $p$-band. The latter is  given by
\be
\Delta_{\text{p}} \left( t,t' \right) = v_{pd}(t) \left( t \right) G^{0}_{\text{p}} \left( t,t' \right)  v_{pd}(t')\;,
\ee
where $G^{0}_{\text{p}} \left( t,t' \right)$ is the local Green's function of the $p$-band (with a semi-elliptic density of states of bandwidth $4$, as shown in Fig.~\ref{fig_spectr_func_vs_om_beta_10_bath}), and  $v_{pd}=E(t)$ is the field-controlled inter-band hybridization. The time-dependent spectral and occupation functions are computed from the retarded and lesser components of the Green's functions via backward Fourier transformation:
\be
\begin{split}
& A_{a} \left( \omega, t \right) = - \frac{1}{\pi} \text{Im} \int_0^{s_\text{max}} d s \ G^{R}_{a} \left( t , t - s \right) e^{-i \omega s} \;, \\
& A^{<}_{a} \left( \omega, t \right) = \frac{1}{\pi} \text{Im} \int_0^{s_\text{max}} d s \ G^{<}_{a} \left( t , t - s \right) e^{-i \omega s} \;.
\end{split}
\ee
The total spectral function is defined as the sum, $A \left( \omega, t \right) = A_{1} \left( \omega, t \right) + A_{2} \left( \omega, t \right)$, and similarly for $A^{<} \left( \omega, t \right)$.

\section{Results} \label{secIII}


\subsection{Time-dependent charge transfer}
\label{secIIIA}

\begin{figure}
 \centering \includegraphics[width=0.5\textwidth]{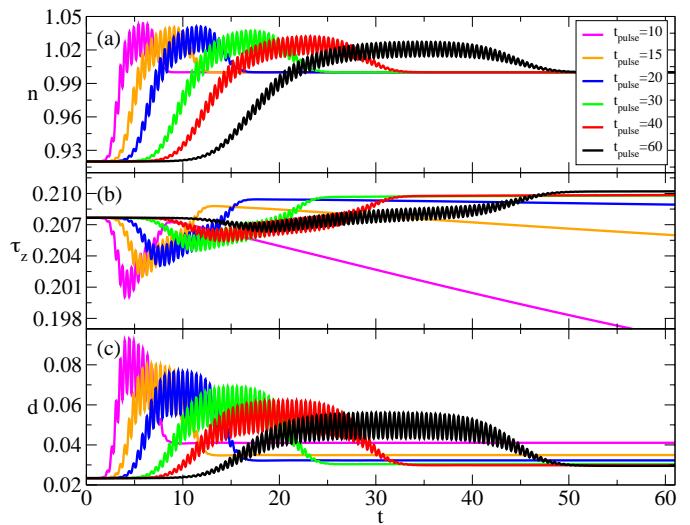}
  \caption{
Time dependence of the occupation per site $n = n_{1} + n_{2}$ (panel (a)), of the orbital imbalance $\tau_{z}$ (panel (b)) and of the double occupation $d$  (panel (c)) for several pulse durations $t_{\text{pulse}}$. Each pulse is tailored to minimize the occupation of the UHBs in the final state.}
  \label{fig_docc_tauz_vs_time_D}
\end{figure}

\begin{figure}
 \centering \includegraphics[width=0.5\textwidth]{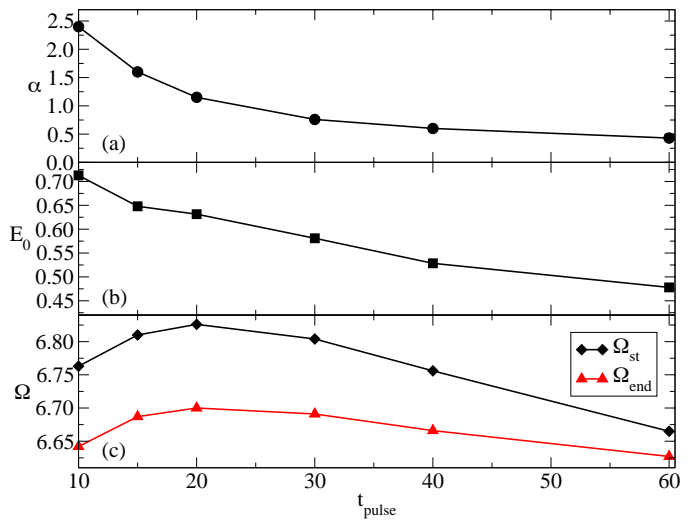}
  \caption{ Optimized pulse parameters $\alpha$ (panel (a)), $E_{0}$ (panel (b)) and $\Omega_{\text{min}}$, $\Omega_{\text{max}}$ (panel (c)) as a function of the pulse duration $t_{\text{pulse}}$, for the pulses used in the simulations shown in Fig.~\ref{fig_docc_tauz_vs_time_D}.}
  \label{fig_pulse_param_vs_tpulse}
\end{figure}

\begin{figure*}
 \centering \includegraphics[width=1\textwidth]{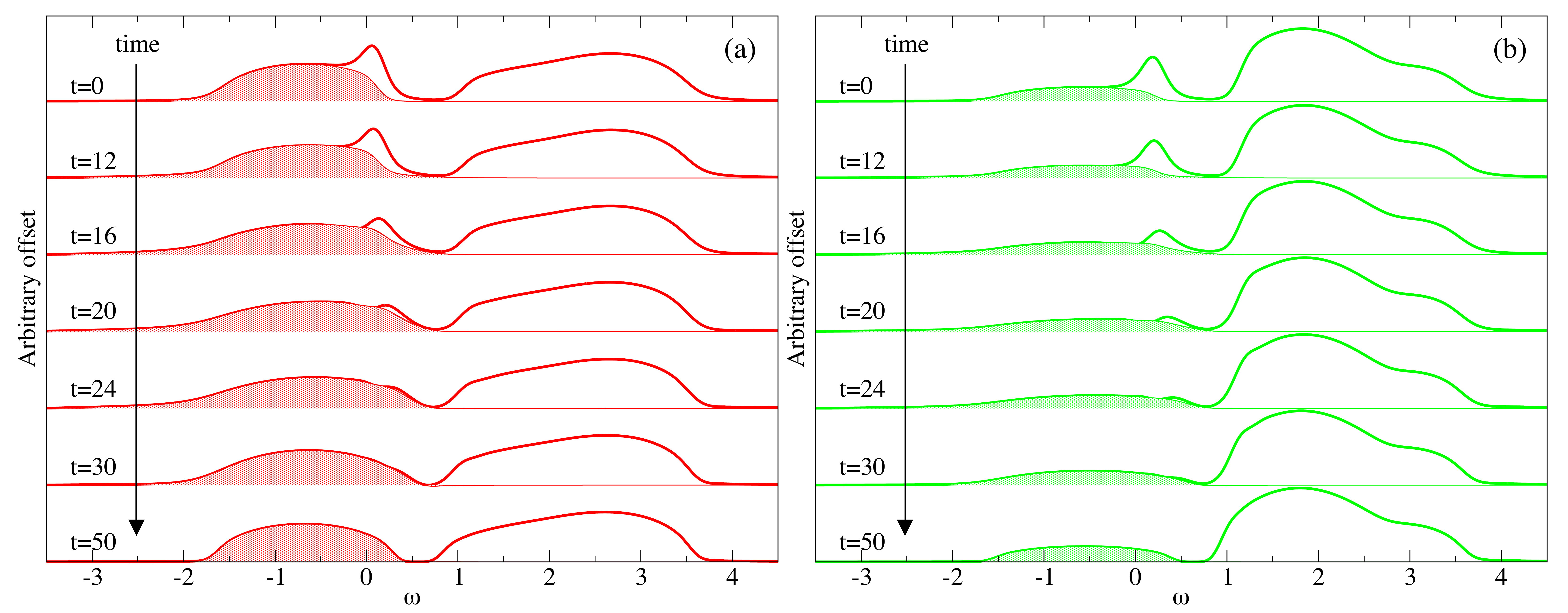}
  \caption{ Panel (a): spectral functions (lines) of band $1$ at different times and respective occupations (shaded areas) during and after the pulse with $t_{\text{pulse}} = 40$ analyzed in \figu{fig_docc_tauz_vs_time_D}. Panel (b): same as for panel (a), but for band $2$. The starting time for the simulations in this figures is $t=-30$, and the pulse arrives at $t=0$. }
  \label{fig_dos_time_evolution}
\end{figure*}

In this section, we present the results obtained by solving the model \equ{tot_ham} using nonequilibrium DMFT, starting from the equilibrium configuration depicted in \figu{fig_spectr_func_vs_om_beta_10_bath}(b). The dipolar coupling of the valence bands with the p-band has the net effect of increasing the filling of the valence bands, see \figu{fig_docc_tauz_vs_time_D}(a). Since our main aim is to induce a metal-to-Mott insulator phase transition in the valence bands, we target the nearest commensurate filling of the valence bands, $n=1$. By applying laser pulses of different durations, we realize that the dynamics of $n$ can be divided essentially into two steps: (i) A growth from the initial value $n \sim 0.92$ to an occupation $n > 1$ larger than the target one, and then (ii), a convergence of $n$ to the desired occupation $n \sim 1$, see \figu{fig_docc_tauz_vs_time_D}(a). The maximum value of the occupation reached during the dynamics is higher for shorter pulses. In passing, we note that each pulse is tailored to minimize the occupation of the UHBs in the final state. The resulting pulse parameters for the pulses used in \figu{fig_docc_tauz_vs_time_D} are shown in Fig.~\ref{fig_pulse_param_vs_tpulse}, while the optimization procedure is described in Sec.~\ref{secIIIB}. 
 
Once the p-band is decoupled from the valence bands, the number of electrons in the valence bands stays fixed because direct charge transfer processes are activated only in the presence of the electric field. Other recombination channels, which are not included in the model, should be weak in the real material and therefore not relevant on the timescale of the present study. In particular, the main recombination channels would be either phonon-assisted, which is slow because of the large energy mismatch between phonons and the p-d energy separation, and p-d electron-electron scattering, which is weak if the p-hole and the photodoped electron are mainly located on different atoms (oxygen and titanium).

The double occupation $d$ shows qualitatively similar dynamics as the density, see \figu{fig_docc_tauz_vs_time_D}(c). The final value of $d$ decreases for longer pulses, suggesting a final state with a smaller number of excitations. Finally, the charge distribution of the model is also characterized by the orbital imbalance $\tau^{z}$, shown in \figu{fig_docc_tauz_vs_time_D}(b). At earlier times, $\tau^{z}$ changes roughly opposite to the  total occupation: When $n$ increases, $\tau^{z}$ decreases and vice versa. After the pulse, the orbital imbalance continues to decrease, in particular after the shorter pulses $t_{\text{pulse}} = 10, 15, 20$, indicating inter valence-band thermalization processes.

To get more insight into what happens during the dynamics, we show in \figu{fig_dos_time_evolution} some snapshots of the spectral functions of the valence bands during one of the longer pulses considered in \figu{fig_docc_tauz_vs_time_D}. The quasiparticle peaks of the two bands observed in equilibrium are gradually suppressed, and entirely vanish once the pulse is over. Thus, while at the initial stage of the dynamics the system is in a correlated metallic state, after the photo-doping the system reaches a Mott state. As seen in \figu{fig_spectr_func_vs_om_beta_10_bath}, the initial metallic state already shows a preformed Mott gap within the unoccupied region of the spectrum. This gap is reduced at intermediate stages of the dynamics due to the hybridization of the valence with the lower energy band (see the intermediate spectral functions shown in \figu{fig_dos_time_evolution}). In the final state, a well-defined Mott gap is restored.


\subsection{Final state and pulse optimization} \label{secIIIB}

\begin{figure}
 \centering \includegraphics[width=0.5\textwidth]{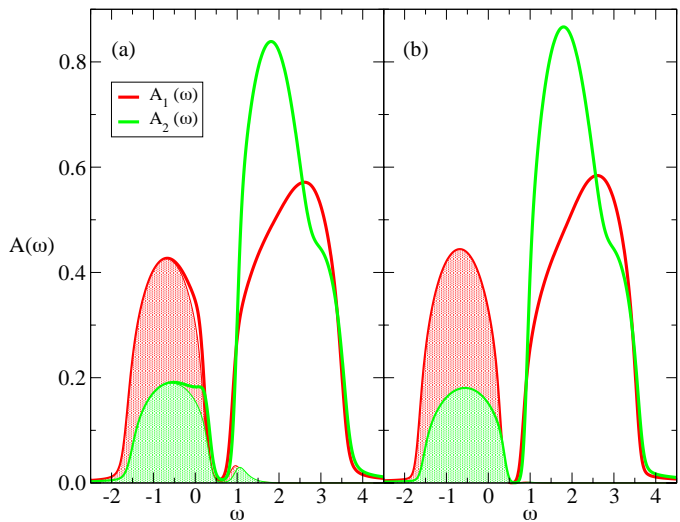}
  \caption{Spectral functions for the two orbitals $A_{1} \left( \omega \right)$ (red) and $A_{2} \left( \omega \right)$ (green), and respective occupations (shaded areas) computed by a backward Fourier transformation at time $t = 130$ after the pulses $t_{\text{pulse}} = 10$ (panel (a)) and $t_{\text{pulse}} = 40$ (panel (b)) analyzed in \figu{fig_docc_tauz_vs_time_D}.}
  \label{fig_spectr_funct_pulse_500_2000_comp}
\end{figure}

Following the original aim of transferring the valence bands into a Mott state, we now analyze the final state after the pulse as a function of the pulse parameters. The insulating character of the state after the pulse strongly depends on the pulse duration $t_{\text{pulse}}$. Figures~\ref{fig_spectr_funct_pulse_500_2000_comp}(a) and (b) show the spectral functions of the Mott insulators reached after pulses of duration $t_{\text{pulse}} = 10$ and $t_{\text{pulse}} = 40$, respectively. It becomes apparent that for the shorter pulse, the final state looks like a Mott insulator with both occupation in the UHB and hole occupation in the lower Hubbard band (LHB), and still some signatures of metallic character such as quasiparticle peaks in particular on the upper edge of the lower Hubbard band. This state can be called a photo-doped Mott-insulator \cite{dasari2020photoinduced}. In contrast, for the longer pulse, the final insulating state resembles a low-temperature Mott insulator.

One could attempt to extract an effective temperature of these states (and also for the transient states) by analyzing the distribution function $F \left( \omega, t \right) = \frac{A^{<} \left( \omega, t \right)}{A \left( \omega, t \right)}$ in terms of a Fermi-Dirac fit with a given inverse temperature $\beta_{\text{eff}}(t)$. This would also tell if an effective temperature is well-defined during the dynamics, i.e., whether the laser induced charge transfer can be viewed as an adiabatic process. However, even for the longer pulses, we find a substantial mismatch between the time-dependent distribution function and the Fermi-Dirac distribution during the pulse. Also after the pulse, a Fermi-Dirac fit is difficult: for shorter pulses, the simultaneous hole occupation in the LHB and electron occupation in the UHB (as in \figu{fig_spectr_funct_pulse_500_2000_comp}a) does not fit a Fermi distribution, while for longer pulses, where the system becomes closer to a low-temperature Mott insulating state, a Fermi-Dirac fit is not possible simply because a distribution cannot be properly defined in the region of the gap and is either zero or unity outside.

\begin{figure}[tbp]
 \centering \includegraphics[width=0.5\textwidth]{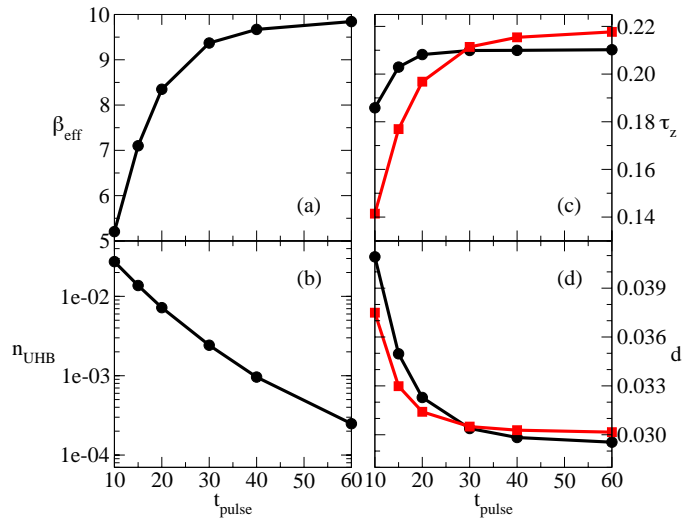}
  \caption{ (a) Effective inverse temperature $\beta_{\text{eff}}$ of the system after the action of a time dependent pulse of duration $t_{\text{pulse}}$. The inverse temperature is computed by looking at the internal energy of the system after the pulse. (b) Occupation of the upper Hubbard band as a function of the duration of the pulse $t_{\text{pulse}}$. Panel (c) and (d): orbital imbalance and double occupation of the system, respectively, at the latest time of our simulations (circles) and in an equilibrium state with the same temperature as in panel (a) (squares).}
    \label{fig_uhb_log_beta_fin_vs_tpulse}
\end{figure}

We therefore proceed by characterizing the final state in terms of its energy. We compare the state of the system with a thermal state sharing the same internal energy and a total valence occupation $n = 1$. The inverse temperature $\beta_{\rm eff}$ of this equilibrium reference state is shown in \figu{fig_uhb_log_beta_fin_vs_tpulse}(a). We can distinguish two main regimes here: When the pulse is short, the effective temperature of the system is increased with respect to the temperature of the initial equilibrium system $\beta = 10$, while with longer pulses there is the possibility to at least keep the effective temperature close to the initial value. Essentially, we have a transition from a regime where the effect of the laser pulse is to produce a photo-doped Mott insulator to a range where the entropy increase in the valence bands remains controlled, as in the cooling by photo-doping protocol. The red squares shown in \figu{fig_uhb_log_beta_fin_vs_tpulse}(c) and (d) represent the values of the orbital imbalance and the double occupation, respectively, as a function of the pulse duration obtained from the equilibrium reference state with the inverse temperatures shown in \figu{fig_uhb_log_beta_fin_vs_tpulse}(a). The black circles in \figu{fig_uhb_log_beta_fin_vs_tpulse}(c) and (d) are instead the values of $\tau_{z}$ and $d$ at the end of the total simulation time for the curves shown in \figu{fig_docc_tauz_vs_time_D}(b) and (c). The crossing of the curves around $t_{\text{pulse}} \sim 30$ separates two regimes: Below $t_{\text{pulse}} \sim 30$, the system looks warmer than the corresponding equilibrium state if we look at the double occupation, while it appears colder if we look at the orbital imbalance; above $t_{\text{pulse}} \sim 30$, the opposite behavior occurs. While the double occupation is essentially constant in the state reached once the pulse is over, see \figu{fig_docc_tauz_vs_time_D}(c), the orbital imbalance evolves towards the corresponding equilibrium value. Thus it keeps decreasing for short pulses (characterized by $t_{\text{pulse}} \le 20$), and slowly increasing for longer pulses, see \figu{fig_docc_tauz_vs_time_D}(b).

\begin{figure}
 \centering \includegraphics[width=0.5\textwidth]{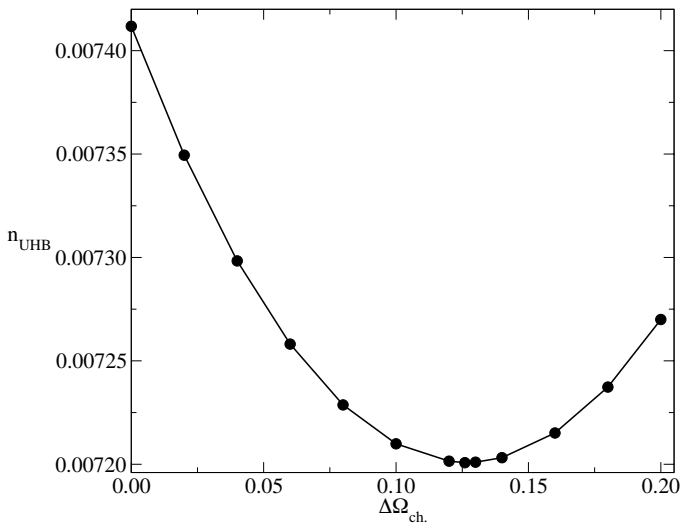}
  \caption{ Occupation of the UHB as a function of the chirping amplitude $\Delta \Omega_{\text{ch.}}$ for pulses with $t_{\text{pulse}} = 20$, $\alpha = 1.15$ and $\Omega_{\text{end}} = 6.700$. }
  \label{fig_nuhb_vs_chirp_ampl}
\end{figure}

As expected, different pulse shapes \equ{pulse} lead to different final states. To choose the specific form of the laser pulse in the simulations so far, we optimized the parameters $\Omega_{\text{end}}$, $\Omega_{\text{st}}$, $\alpha$, and $E_0$ for each pulse duration $t_{\text{pulse}}$, keeping the final occupation $n=1$ fixed. For the optimization, we start using \equ{pulse} for a non-chirped pulse close to $\Omega_{\text{end}} = \Omega_{\text{st}} = \Omega \sim 6.65$ (set by the p-d splitting) and $\alpha \sim 1$, and determine $E_{0}$ to impose the desired total occupation $n \sim 1$. Considering several values of $\Omega$, we find the configuration that minimizes the occupation of the UHB at given $\alpha$. Repeating this procedure for several values $\alpha$, we optimize the number of excitations in the system for this parameter too. Once we are in the best possible state according to our requirements, we allow $\Omega_{\text{end}}$ to be different from $\Omega_{\text{st}}$, permitting us to optimize the excitations number respect to the unchirped configuration slightly. (Certainly an automatic multi-parameter pulse optimization based on machine learning techniques would be intriguing, but computationally costly due to the potentially large number of simulations required in the optimization.) The optimized parameters are shown in \figu{fig_pulse_param_vs_tpulse}. The optimal pulse shape is indeed found to be slightly chirped, with $\Omega_{\text{end}}$ smaller than $\Omega_{\text{st}}$, and a chirping amplitude $\Delta \Omega_{\text{ch.}} = \Omega_{\text{st}} - \Omega_{\text{end}}$ comparable to the gap size, see \figu{fig_pulse_param_vs_tpulse}(c). Figure.~\ref{fig_nuhb_vs_chirp_ampl} shows the occupation of the UHB $n_{\text{UHB}}$ against $\Delta \Omega_{\text{ch.}}$ for pulses characterized by $t_{\text{pulse}} = 20$, $\alpha = 1.15$ and $\Omega_{\text{end}} = 6.700$. Each point in \figu{fig_nuhb_vs_chirp_ampl} is computed at a different amplitude $E_0$ to impose the total occupation of the final state $n \sim 1$. The minimum of $n_{\text{UHB}}$ occurs at $\Delta \Omega_{\text{ch.}} = 0.126$, corresponding to $\Omega_{\text{st}} = 6.826$, as reported in \figu{fig_pulse_param_vs_tpulse}(c).

A final question we would like to address is at which speed the transition from the metallic to the Mott insulating regime can be achieved. Since we target to keep a small number of excitations, we plot the number of electrons in the UHB once the pulse is over as a function of the pulse duration, as shown in \figu{fig_uhb_log_beta_fin_vs_tpulse}(b). Not surprisingly, the shorter the pulse, the higher the number of excitations in the system. We expect the curve to follow the exponential decay $C e^{-E_{g}^{\text{fit}} t_{\text{pulse}}}$, where $E_{g}^{\text{fit}}$ should be of the order of the bandgap $E_{g}$ for the best possible pulse process. By fitting the curve \figu{fig_uhb_log_beta_fin_vs_tpulse}(b) with the exponential function, we obtain $C \sim 0.102$ and $E_{g}^{\text{fit}} \sim 0.132$, with the second parameter of the same order of magnitude of the gap size $E_{g} \sim 0.2$. We notice that, while the three points at $t_{\text{pulse}} \le 20$ in \figu{fig_uhb_log_beta_fin_vs_tpulse}(b) are well fitted, the last three depart from the fitting line. For this, one also must keep in mind that the exponentially small excitation densities become increasingly difficult to measure numerically. The decrease in $n_{\text{UHB}}$ observed in \figu{fig_uhb_log_beta_fin_vs_tpulse}(b) over several order of magnitudes indicates what should be the approximate pulse length to get the desired number of excitations in the system.

\section{Conclusions}\label{secIV}
In this study, we analyzed the possibility to induce an ultrafast metal-to-valence Mott insulator transition in a model system relevant to the physics of the oxygen-enriched transition metal compound LaTiO$_{3 + \text{x}}$. Through non-equilibrium dynamical mean-field theory (with the non-crossing approximation as an impurity solver), we considered the effect of the dipolar photo-doping of electrons from a core state representative of the broad p-derived oxygen band to the valence states that mimic the $\text{t}_{2 \text{g}}$ manifold at the Fermi level. An external laser pulse opens a charge transfer channel from the oxygen related bands to the valence bands. At each given pulse duration, we optimized the shape of the pulse to reach the desired final occupation and simultaneously minimize the number of excitations in the upper Hubbard bands. Depending on the pulse length, we get a crossover from a regime where the final state is a Mott insulator characterized by an effective temperature almost equal to the one of the initial correlated metallic system to a regime where this effective temperature is substantially larger. The crossover is at pulse durations as short as $30$ hopping times.

Although the model under investigation is highly simplified, this suggests that a photo-induced charge transfer to a valence Mott-insulator can be realized in LaTiO$_{3 + \text{x}}$ on the $100 \ \mathtt{fs}$ timescale without substantially heating the system. As in the cooling by photo-doping mechanism, the limited heating can be related to the fact that a large amount of entropy is stored in the oxygen related bands after the photo-induced charge transfer.

\section*{Acknowledgements}
We thank Philipp Werner for useful discussions. This work was supported by the ERC starting grant No. 716648. The authors gratefully acknowledge the computational resources and support provided by the Erlangen Regional Computing Center (RRZE).


%

\end{document}